\documentclass[aps,prl,reprint,superscriptaddress,nobibnotes]{revtex4-2}

\usepackage{amsmath,amssymb,graphicx,color,verbatim,soul,braket}
\usepackage[english]{babel}
\makeatletter
\adddialect\l@en\l@english
\makeatother
\usepackage{scalerel,stackengine}
\usepackage{upgreek}
\usepackage{bm}
\usepackage[hidelinks]{hyperref}

\newcounter{parlabel}
\renewcommand{\theparlabel}{\Alph{parlabel}}
\newcommand{\parlabel}[1]{
  \refstepcounter{parlabel}
  \noindent\textbf{\theparlabel.} \label{#1}
}

\begin{document}

\title{Fast Coherent Splitting of Bose-Einstein Condensates}

\author{Yevhenii Kuriatnikov}
\affiliation{Vienna Center for Quantum Science and Technology, Atominsitut, TU Wien, Stadionallee 2, 1020 Vienna, Austria}

\author{Nikolaus W\"urkner}
\affiliation{Automation and Control Institute, TU Wien, Gußhausstraße 27-29, 1040 Vienna, Austria}

\author{Karthikeyan Kumaran}
\affiliation{Vienna Center for Quantum Science and Technology, Atominsitut, TU Wien, Stadionallee 2, 1020 Vienna, Austria}

\author{Tiantian Zhang}
\affiliation{Vienna Center for Quantum Science and Technology, Atominsitut, TU Wien, Stadionallee 2, 1020 Vienna, Austria}

\author{M. Venkat Ramana}
\affiliation{Vienna Center for Quantum Science and Technology, Atominsitut, TU Wien, Stadionallee 2, 1020 Vienna, Austria}

\author{Andreas Kugi}
\affiliation{Automation and Control Institute, TU Wien, Gußhausstraße 27-29, 1040 Vienna, Austria}
\affiliation{AIT Austrian Institute of Technology, Giefinggasse 4, 1210 Vienna, Austria}

\author{J\"{o}rg Schmiedmayer}
\affiliation{Vienna Center for Quantum Science and Technology, Atominsitut, TU Wien, Stadionallee 2, 1020 Vienna, Austria}

\author{Andreas Deutschmann-Olek}
\email{Deutschmann@acin.tuwien.ac.at}
\affiliation{Automation and Control Institute, TU Wien, Gußhausstraße 27-29, 1040 Vienna, Austria}

\author{Maximilian Pr\"ufer}
\email{maximilian.pruefer@tuwien.ac.at}
\affiliation{Vienna Center for Quantum Science and Technology, Atominsitut, TU Wien, Stadionallee 2, 1020 Vienna, Austria}

\begin{abstract}

Preparation of non-trivial quantum states without introducing unwanted excitations or decoherence remains a central challenge in utilizing ultracold atomic systems for quantum simulation. We employ optimal control methods to realize fast, coherent splitting of a one-dimensional Bose-Einstein condensate, achieving minimal classical excitations while preserving quantum correlations. Furthermore, we explore two-step protocols in which controlled classical motion is first induced and subsequently suppressed via tailored control sequences. Our experiments highlight  the potential of optimal control for quantum state engineering and dynamical control in many-body quantum systems.

\end{abstract}

\maketitle

Ultracold atomic systems provide a versatile platform for the implementation of quantum simulators
{\cite{bloch_quantum_2012,PRXQuantum.2.017003}},
enabling the investigation of complex quantum many-body dynamics with unprecedented control and tunability. 
In many applications, it is essential to steer the system into nontrivial quantum states.
In the presence of interactions, it may be hard to manually design control sequences, especially when high speed is required or certain other boundary conditions have to be met.

The use of optimal control ($OC$) has emerged as a powerful methodology for this purpose, offering fast, time-efficient, and robust pathways for quantum state engineering \cite{Koch2022QuantumControl,RevModPhys.91.045001}.  
$OC$ requires good knowledge of the system under study and an accurate mathematical model of its dynamical processes, which can be challenging to obtain. Quantum optimal control 
not only facilitates fast state preparation  \cite{omran2019generation}  but also accelerates transport tasks \cite{PhysRevX.11.011035} and can suppress undesired excitations \cite{PhysRevA.75.023602} and decoherence, which are otherwise limiting factors in quantum technology applications. 

In this letter, we demonstrate fast, coherent splitting of a one-dimensional Bose-Einstein condensate using $OC$. We introduce an effective model used for the $OC$ protocols and discuss an efficient calibration method. Our results demonstrate that we are able to reduce classical excitations to a minimum for a wide range of splitting speeds. 
We achieve strikingly fast splitting, approaching the system's theoretical speed limit, while preserving quantum correlations, offering a pathway for manipulating the quantum properties of the two condensates.

\begin{figure}[t!]
    \centering
    \includegraphics[width=0.9\linewidth]{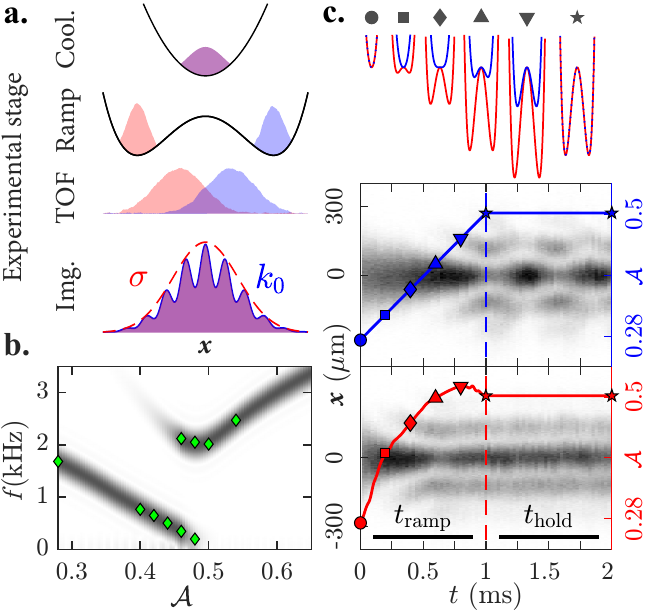}
\caption{
\textbf{ Experimental scheme}  
\textbf{a.} A BEC is prepared in a single-well harmonic trap. It is split into a double-well during \( t_{\text{ramp}} \) using RF-dressing; to probe the resulting state, it is held for \( t_{\text{hold}} \), released from trap, and imaged after a TOF of 43.5\,ms. Fluorescence imaging captures interference fringes, from which the transverse cloud size \( \sigma \) and fringe spatial frequency \( k_0 \) are extracted.  
\textbf{b.} Trap frequencies (green diamonds) as a function of RF-dressing amplitude \(\mathcal{A} \) determined from oscillatory motion of the displaced BEC are compared with frequencies extracted from simulations using the calibrated reduced model.  
\textbf{c.} Comparison of linear (blue) and optimal-control (red) ramps. The evolution of the interference pattern during ramp and hold times is shown alongside the simulated potential at selected times. Linear ramps induce strong excitations, whereas $OC$ ramps achieve splitting with negligible excitation.
}
    \label{fig:1_setup}
\end{figure}

\textit{{Coherent splitting for quantum state preparation.---}}
It has been demonstrated that the splitting of Bose–Einstein condensates (BECs) \cite{Schumm2005,Barker_2020} 
into double-well potentials enables quantum state preparation protocols \cite{Esteve2008}. Interferometric and metrological applications have particularly benefited from squeezed and entangled states generated in split BECs \cite{pezze_quantum_2018},
enabling enhanced sensitivity beyond the standard quantum limit (SQL).   
The quality of such processes depends critically on the suppression of motional excitations and the preservation of coherence and quantum correlations. Theoretical and experimental studies have shown that fast, coherent manipulations of ultracold atoms in state-dependent potentials or anharmonic traps are possible via optimal control \cite{vanFrank2014,PhysRevA.74.022312}.

We implement our experiments with a one-dimensional BEC of $^{87}$Rb in a magnetic trap on an atomchip~\cite{Folman_AtomChip,reichel2011atom}. Radio-frequency (RF) dressing \cite{hofferberth2006radiofrequency,Lesanovsky2006,Lesanovsky2006a} enables continuous and reproducible tuning of the trapping potential from a single-well configuration to a double-well geometry using a single parameter: the amplitude $\mathbf{\mathcal{A}}$ of the RF current. 
Detection is performed by acquiring a fluorescence image after a time-of-flight (TOF) of \(43.5 \,\mathrm{ms}\)~\cite{Buecker_2009}. 
In earlier work, it has been demonstrated that this type of control and readout allows us to detect number squeezing and entanglement \cite{berrada2013integrated,zhang2024squeezing}, but tunability of quantum properties was not possible.

\textit{OC model.---}
Precise feed-forward control relies on mathematical models that accurately capture the dynamics of the physical system while remaining simple enough to enable reliable calibration and tuning.
To design optimizable and experimentally implementable control protocols for BEC splitting, we seek reduced models that describe the essential dynamics under time-dependent control. In our case, this requires not only an accurate representation of the trapping potential but also a framework for modeling the real-time evolution of the system.

For the latter, we focus on the evolution in the transverse (splitting) direction; the longitudinal direction remains approximately static during typical experimental time scales. The dynamical evolution of the classical nonlinear motion is modeled by a reduced Gross-Pitaevskii equation (GPE)
\begin{align*}
    i\hbar\frac{\partial\Psi}{\partial t} = \left(-\frac{\hbar^2}{2m}\frac{\partial^2}{\partial x^2} + V(x, \mathcal{A}(t)) + g_{\perp}|\Psi|^2\right)\Psi,
\end{align*}
where the time-dependent effective potential \( V(x, \mathcal{A}(t)) \) and the effective transverse interaction strength \( g_{\perp} \) are open parameters. Accurate calibration of these parameters is essential for reproducing the experimental dynamics and ensuring the success of optimal control protocols.
 
For the effective potential \( V(x, \mathcal{A}) \), we employ the most simple double-well model \cite{Lesanovsky2006,Lesanovsky2006a}
\[
V(x, \mathcal{A}) = a_2(\mathcal{A}) x^2 + a_4(\mathcal{A}) x^4,
\]
where the coefficients are parameterized to reflect the curvature and separation of the wells as a function of the control parameter $\mathcal{A}$. Strikingly, this reduced model captures all relevant physical features of the system with sufficient accuracy for optimal control purposes. 

The parameters of the effective model are calibrated from experimental observations. This is achieved by comparing the dynamical evolution predicted by the model with experimental observations and adjusting the model parameters to ensure agreement (see Fig.~\ref{fig:1_setup}b). A more detailed outline of the calibration procedure can be found in the Appendix \ref{par:C}, \ref{par:D},  and \ref{par:E}, and is discussed in full detail in \cite{Wurkner2025}. 

In Fig.~\ref{fig:1_setup}c, we present a representative comparison of the performance of linear and $OC$ splitting protocols. The figure shows the respective control ramps, the evolution of the interference fringe patterns during the ramp and subsequent hold time, as well as the corresponding time-dependent shape of the double-well potential throughout the splitting process. The superior performance of the $OC$ protocol is evident already from the fringe evolution, as the fringe pattern remains static after the end of the $OC$ ramp.

\textit{Shortcut to adiabatic splitting.---}
\begin{figure*}[t]
    \centering
    \includegraphics[width=\linewidth]{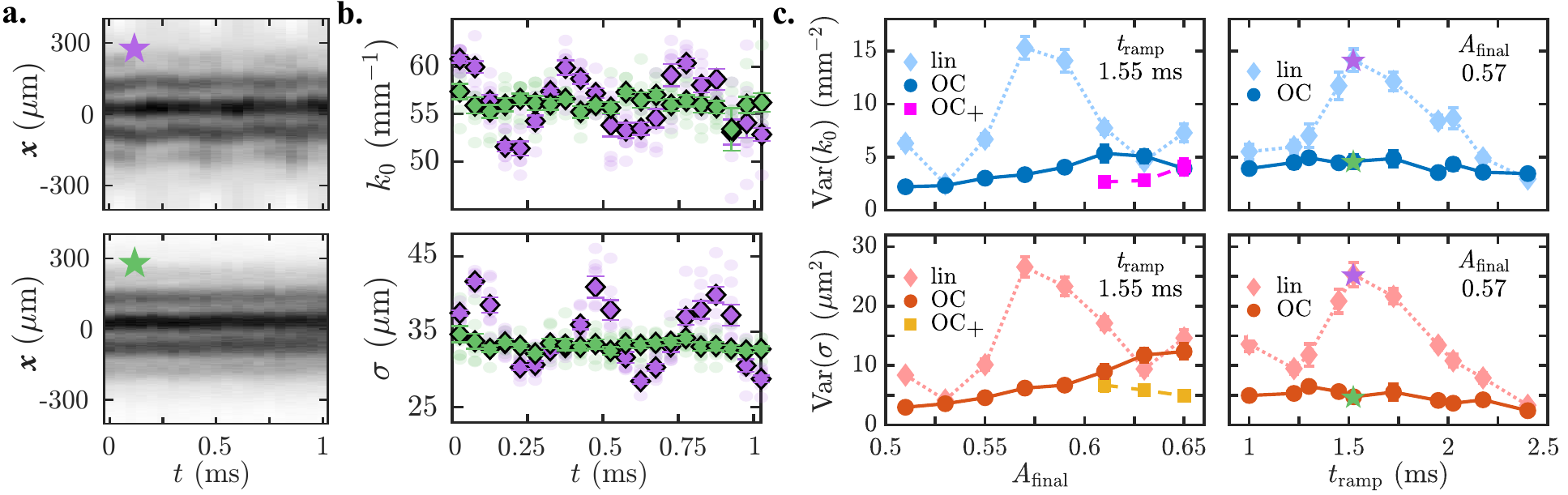}
\caption{
\textbf{$\mathbf{OC}$ minimization of splitting-induced excitations.} 
\textbf{Panel a.}: We show averaged over multiple realizations fringe profiles taken after the end of the respective splitting ramps. 
\textbf{Panel b.} Displayed are the extracted fringe spacing $k_{0}$ and transversal width $\sigma$ for both linear (magenta) and $OC$ (green) ramps (cf. \textbf{a.}). Diamonds represent mean values, while transparent circles are single shots. This analysis highlights the suppression of sloshing dynamics achieved with $OC$ protocols compared to linear ramps.
\textbf{Panel c. Top row:} Variance of the fringe spacing $k_{0}$ during \( t_{\text{hold}} \) after splitting (sloshing).  
\textbf{Panel c. Bottom row:} Variance of the transverse width \( \sigma \) (breathing).  
\textbf{Panel c. Left column:} Fixed ramp duration \( t_{\text{ramp}} \), varying final splitting strength \( \mathcal{A}_{\text{final}} \).  
\textbf{Panel c. Right column:} Fixed final splitting \( \mathcal{A}_{\text{final}} \), varying ramp duration. Two points marked with stars refer to \textbf{a.}
Linear ramps (diamonds) show strong excitations with non-trivial dependence on both ramp duration and final splitting. Optimal control ramps (circles) consistently suppress excitations across a wide parameter range. At larger \( \mathcal{A}_{\text{final}} \), the performance of $OC$ deteriorates due to increased sensitivity to modeling inaccuracies. We introduce a modified control scheme, \( \mathrm{OC}_{+} \) (squares), employing rescaled ramp durations by a few percent, which restores performance and further suppresses excitations.
}
    \label{fig:2_resonance}
\end{figure*}
Performing splitting via non-adiabatic, e.\,g., fast linear ramps, excites classical collective modes. The most prominent of these are oscillations of the inter-well distance $d$, referred to as \textit{sloshing}, and oscillations of the transverse width of the condensates $\tilde{\sigma}$, known as \textit{breathing}  
\cite{RevModPhys.71.463}.
We study resulting post-TOF interference patterns, from which we extract the fringe spatial frequency (fringe spacing) \(k_0\) and the transverse width \(\sigma\) (cf. Fig. \ref{fig:1_setup}a), which relate, respectively, to the inter-well separation and the in-situ spatial extent of the individual condensates (see Appendix \ref{par:A}).

As an example, Fig. \ref{fig:2_resonance}a shows the evolution of the fringe pattern over a $1\,\text{ms}$ hold time for both linear (top) and $OC$ (bottom) ramps. In Fig. \ref{fig:2_resonance}b, we present the extracted fringe spatial frequency (fringe spacing) $k_0$ and the transverse width $\sigma$, displaying individual experimental realizations along with their mean values. 
The data demonstrate clear suppression of excitations in post-ramp dynamical evolution for the $OC$ ramp. 
For quantitative analysis of the amplitude of these excitations without assuming a single dominant frequency, we evaluate two observables over the hold time: the variance of $k_0$ and the variance of $\sigma$ (see Appendix \ref{par:B}).

In Fig.\ref{fig:2_resonance}c, we show the performance of the $OC$ splittings for two different scenarios:  (I) varying the final splitting amplitude $\mathbf{\mathcal{A}_{\text{final}}}$ while fixing the ramp duration $t_{\text{ramp}}$ (first column in Fig.\ref{fig:2_resonance}c), and (II) varying the ramp duration $t_{\text{ramp}}$ while fixing the final amplitude $\mathbf{\mathcal{A}_{\text{final}}}$ (second column in Fig.\ref{fig:2_resonance}c). For comparison, we show the results obtained using a linear ramp with identical parameters. Notably, the linear ramp exhibits non-trivial behavior, revealing a resonance-like structure in the excitation amplitude. Strikingly, the $OC$ outperforms the linear ramps for almost all scenarios. For larger final dressing amplitudes, the performance is limited by the simplicity of the model. 
However, with a simple linear time rescaling of the trajectory from \( \mathcal{A}(t)  \) to \( \mathcal{A}((1-\gamma) t)  \) (labeled by $OC_+$), with \( \gamma \ll 1 \),  we achieve superior performance.

The presented $OC$ scheme allows us to perform splitting procedures for a broad range of timescales without exciting classical motion. In particular, the $OC$ approach allows access to significantly shorter timescales compared to linear ramps (which typically require durations of $10\, \text{ms}$ or longer to avoid excitations), thereby opening the door to splitting times that are comparable or even faster than interaction time scales.

\textit{OC engineering.---}
\begin{figure}[t]
    \centering
    \includegraphics[width=0.98 \linewidth]{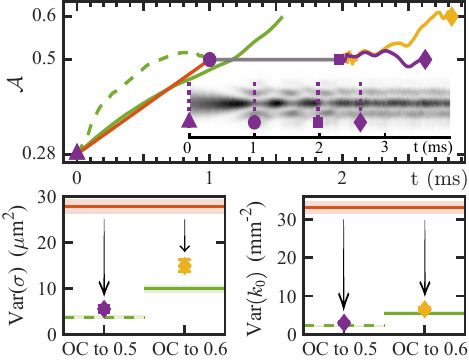}
\caption{
\textbf{Optimal control engineering.}  
Dressing amplitude \(\mathcal{A}\) as a function of time for different engineering scenarios. We first perform a fast linear splitting (red line) into a coupled double-well trap, introducing strong motional excitations. After a hold time of 1 ms (gray segment), we apply an $OC$ sequence to either (I) further split into a more decoupled trap (orange trajectory) or (II) remove excitations while staying in the same coupled configuration (purple trajectory).
Inset shows TOF interference fringe images with annotated markers: start of initial splitting (triangle), end of linear ramp (circle), start of $OC$ sequence (square), end of $OC$ sequence (diamond).
The bottom panels show the variance of transverse cloud width \( \sigma \) and fringe spacing \( k_0 \), comparing trajectories with standard singular linear (red line) and $OC$ (green lines) ramps. Results demonstrate that excitation-minimizing $OC$ sequences can reach near-optimal performance even when starting from highly non-adiabatic initial conditions.}
    \label{fig:3_oce}
\end{figure}
So far, we have demonstrated control protocols capable of mitigating motion induced during the splitting process. However, residual sloshing can modulate key parameters such as the tunnel coupling, thereby affecting the properties of the final state~\cite{PhysRevB.106.075426, PhysRevLett.129.080402, zhang2024squeezing}. Hence, achieving control over the motional degrees of freedom is crucial for creating engineered excitations to manipulate the resulting quantum state.

In Fig.~\ref{fig:3_oce}, we present experiments demonstrating that an initially excited motion can be suppressed by applying a secondary control ramp. We deliberately generate strong motion by a quick linear ramp to a strongly coupled trap $\mathbf{\mathcal{A}}=0.5$. After a $1 \, \text{ms}$ hold time, we apply one of two $OC$ sequences: either to end up again in the coupled double-well with $\mathbf{\mathcal{A}}=0.5$ or to ramp up to a decoupled double-well with $\mathbf{\mathcal{A}}=0.6$ with the goal of suppressing the created excitations. The $OC$ ramps for those scenarios are calculated by simulating the full dynamical evolution of the protocol. Due to small model errors, we had to adjust the hold time before applying the $OC$ ramp to match the simulation and experiment evolution \cite{Wurkner2025}.

As shown in Fig.~\ref{fig:3_oce}, both $OC$ protocols begin with excitations introduced by the initial linear ramp (indicated in red) and subsequently suppress these excitations effectively. The resulting final states exhibit excitation levels comparable to those achieved by a single $OC$ splitting ramp directly targeting the final trap configuration (indicated in green). This indicates that our $OC$ approach in the non-equilibrium scenario, where additional motion is excited, features the same level of control as in the shortcut-to-adiabaticity case.

\textit{Splitting-speed limit.---}
Naturally, the system imposes constraints on the minimum duration over which the splitting can be performed, as instantaneous splitting is physically unattainable. We estimate this lower bound in our setup from numerical simulations. Starting from a single-well harmonic potential, we instantaneously switch to an inverted parabolic potential and evaluate the time required for the atomic density to reach the positions corresponding to the final well locations (without requiring atoms to come to rest there). These simulations give us an estimate of $t_{\text{lim}} \sim 200 \, \mu\text{s}$ for the case of a final coupled trap characterized by $\mathbf{\mathcal{A}_{\text{final}}} = 0.5$, with a corresponding trap frequency of $2 \, \text{kHz}$.

Experimentally, we find that linear ramps with durations shorter than $1 \, \text{ms}$ fail to produce well-defined interference fringes; at the end of such ramps, the BECs are not yet localized in the two wells. In contrast, our $OC$ protocol enables clean and controlled splitting within ramp durations as short as $350 \, \mu\text{s}$.

\textit{Tuning quantum properties.---}
\begin{figure}[t]
    \centering
    \includegraphics[width=0.98 \linewidth]{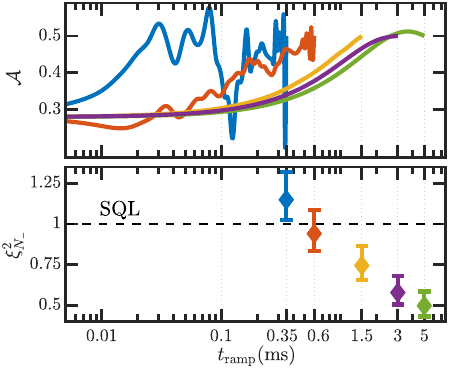}
\caption{\textbf{Squeezing for fast $OC$ ramps. } We perform fast $OC$ ramps to A = 0.5 trap with different ramp times (solid lines, upper panel). In the lower panel, we show the inferred number-squeezing factor. Notably, decreasing the ramp duration reduces the squeezing toward the standard quantum limit (SQL), which is expected since interactions become negligible simply because the evolution time is too short for them to take effect. Even for the fastest ramps, quantum properties are preserved, i.\,e., the relative phase is well-defined. 
}
    \label{fig:4_squeezing}
\end{figure}
During splitting, quantum correlations are created in the relative degrees of freedom of the double well. In the ground state, the interplay of repulsive interactions and tunneling leads to relative atom number squeezing. While the $OC$ protocol provides shortcuts to adiabaticity for the motional degrees of freedom within one well, the different splitting speeds change the impact of interactions. For the limiting case of an infinitely fast ramp, interactions can be neglected, and the splitting could be modeled as a non-interacting beam splitter. Thus, together with the adiabatic preparation of classical motion, tunable splitting times should allow modification of quantum properties. 

We now explore whether quantum features are preserved under $OC$ splitting protocols.
For this, we extract the relative atom number $N_- = N_L - N_R$, where $N_{L|R}$ denote the atom numbers in the two wells. From our experimental data, we infer the number-squeezing factor $\xi^2_{N_{-}} = {\Delta^2 N_-}/{N}$, where \( \Delta^2 \)  denotes the statistical variance, and  $N = N_L + N_R$ is the total atom number. Number squeezing is signaled by $\xi^2_{N_{-}} < 1$, which indicates suppressed number fluctuations below the standard quantum limit (SQL); together with phase coherence, it allows for witnessing entanglement \cite{sorensen_many-particle_2001, pezze_quantum_2018}. 

In Fig.~\ref{fig:4_squeezing}, we investigate how the use of $OC$ ramps enables systematic exploration of the fast splitting regime, where linear ramps fail. By tuning the ramp duration, we study the resulting evolution of the squeezing factor as a function of ramp time. While squeezing remains below the SQL for fast ramps, its magnitude decreases as the ramp duration gets shorter. This behavior is consistent with reduced correlation buildup time in the strongly non-adiabatic regime.

In our system, the detection of coherent state fluctuations is so far hampered by detection noise and technical fluctuations. However, it is important to note that the phase-squeezing factor $\xi^2_{\varphi}$ is decreased for shorter ramp times. This further indicates that quantum coherence persists despite rapid manipulations.

\textit{Conclusion and outlook.---}
In this work, we have demonstrated the application of optimal control engineering to the splitting of a 1D BEC into a double-well potential. Using a calibrated model for the experimental trapping potential, we generated energy-optimal control ramps that significantly suppress excitations such as sloshing and breathing.

We showed that the optimal control ($OC$) approach not only outperforms simple linear ramps across a broad parameter space but also remains effective at timescales where linear ramps cannot even localize atoms in two wells. Furthermore, we demonstrated the flexibility of $OC$ by applying it to complex scenarios, where it was used to remove excitations induced by prior manipulations. This provides a toolbox of building blocks for controlling quantum properties.

We provide a new strategy for optimal control engineering as a powerful tool for quantum state preparation in cold atom systems. The demonstrated preservation of quantum correlations during fast, optimal-control-based splitting provides a tunable approach to initializing low-entropy entangled states, which are essential for quantum-enhanced metrology and form a building block for many-body quantum simulation platforms.

\begin{acknowledgments}
\textit{Acknowledgments.---} This work is supported by the European Research Council: ERC-AdG {\em Emergence in Quantum Physics} (EmQ) under Grant Agreement No. 101097858 and the DFG/FWF CRC 1225 'ISOQUANT', (Austrian Science Fund (FWF) I6949; DOI:10.55776/I6949). Furthermore, we acknowledge support by the Austrian Science Fund (FWF) [10.55776/COE1, P36236-N] and the European Union – NextGenerationEU.
M.P. has received funding from Austrian Science Fund (FWF) [10.55776/ESP396; QuOntM]. 
\end{acknowledgments}
\bibliography{main}

\appendix

\onecolumngrid
\vspace{1em}
\begin{center}
    \textbf{\large Appendix}
\end{center}
\vspace{1em}
\twocolumngrid

\smallskip
{\textit{Appendix }} \textit{\parlabel{par:A}} {\textit{{Relative phase and atom number imbalance.}}} The measurement performed in the experiment is a fluorescence image acquired after $43.5 \, \text{ms}$ of ballistic expansion of the condensate following its release from the trap \cite{Buecker_2009}. 

This image reveals an interference pattern formed by the two overlapping condensates. 
From the fringe structure, we extract key parameters, including the transverse width, relative phase, and fringe spacing; the fringe spacing is related to the inter-well distance. 
Being deep in the 1D regime, the interference fringes are fit using a Gaussian envelope modulated by a cosine function
\begin{equation*}
f(x) \propto \exp\left( -\frac{(x - x_0)^2}{2\sigma^2} \right) \left[ 1 + C \cos(k_0 x + \phi) \right],
\end{equation*}
where $x_0$ is the center of the envelope, $\sigma$ is the envelope width in the time-of-flight image, $C$ is the fringe visibility (contrast), $\phi$ is the fringe phase, and $k_0$ is the spatial frequency, with its corresponding fringe spacing $\lambda = 2\pi/k_0$.

For measuring atom number imbalance, a nearly instantaneous momentum kick is applied just before release, resulting in spatial separation of the two clouds. The counts in the two resulting clouds are integrated to obtain the global atom number in the two wells.

\smallskip
{\textit{Appendix }} \textit{\parlabel{par:B}} {\textit{{Measures for motional excitations }}}
To quantify the amount of breathing and sloshing, we calculate the variances of the transversal width of the cloud $\sigma$ and the spatial frequency $k_0$, respectively. Explicitly, we compute the population variances using
\[
\text{Var}(y_\eta) = \frac{1}{\mathcal{N}_\eta-1} \sum_{i=1}^{\mathcal{N}_\eta} \left( y_{\eta,i} - \bar{y}_{\eta} \right)^2\,,
\]
where \( y_\eta \in \{\sigma, k_{0}\} \) denotes transversal width or fringe spatial frequency, \( y_{\eta,i} \) are the individual data points, and \( \mathcal{N}_\eta \) is the total number of points in the corresponding dataset. The sample means are given by
\[
\bar{y}_{\eta} = \frac{1}{\mathcal{N}_\eta} \sum_{i=1}^{\mathcal{N}_\eta} y_{\eta,i}\,.
\]

To estimate the statistical uncertainty of the variances, jackknife resampling is used \cite{Shao1995JackknifeBootstrap}.

\smallskip
{\textit{Appendix }} \textit{\parlabel{par:C}} {\textit{{The model. }}} 
The model contains five free parameters:

\begin{itemize}
  \item \(\mathcal{A}_s\): the splitting point of the double-well, marking the transition from a single-well to a double-well configuration;
  
  \item \(c\): a scaling factor for the position of the trap minima as a function of \(\mathcal{A}\), described by
  \[
  x_m(\mathcal{A}) = c\, \mathrm{Re}\left[\sqrt{\mathcal{A} - \mathcal{A}_s}\right];
  \]
  
  \item \(\kappa_1\), \(\kappa_2\): scaling factors for the trap curvature at the minima in the single- and double-well regimes, respectively. The effective curvature is given by
  \[
  \frac{\partial^2 V}{\partial x^2}\big(x_m(\mathcal{A}),\mathcal{A}\big)=
  \begin{cases}
  \kappa_1 (\mathcal{A} - \mathcal{A}_s), & \text{for } \mathcal{A} < \mathcal{A}_s, \\
  \kappa_2 (\mathcal{A} - \mathcal{A}_s), & \text{for } \mathcal{A} > \mathcal{A}_s;
  \end{cases}
  \]
  
  \item \(g_{\perp}\): the effective transverse interaction strength.
\end{itemize}

These parameters collectively define the effective one-dimensional potential \( V(x, \mathcal{A}) \) used in the reduced Gross--Pitaevskii model.

\begin{figure}[t!]
    \centering
    \includegraphics[width=0.95\linewidth]{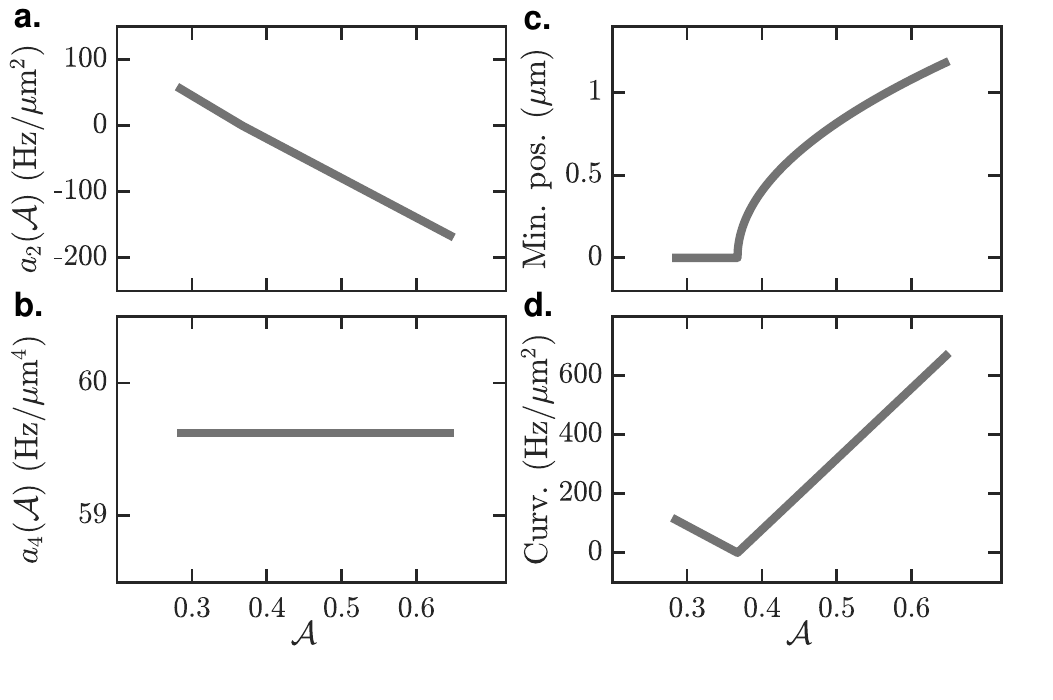}
\caption{\textbf{Parameters of the trap model. } This figure presents the behavior of the model used to reconstruct the trapping potential. \textbf{a.} Coefficient $a_2(\mathcal{A})$ as a function of the splitting parameter $\mathcal{A}$. \textbf{b.} Coefficient $a_4(\mathcal{A})$ characterizes the anharmonic contribution. \textbf{c.} Position of the potential minima \(x_m(\mathcal{A})\) as a function of $\mathcal{A}$. \textbf{d.} Local curvature of the potential \(\kappa_m(\mathcal{A})\) evaluated at the minima \(x_m(\mathcal{A})\).
}
    \label{fig:5_mofel}
\end{figure}

The model parameters are subject to certain constraints, which are supported by full numerical simulation of the trapping potentials for the realistic experimental system \cite{Lesanovsky2006,Lesanovsky2006a}. The quadratic coefficient is assumed to vary linearly with the splitting parameter, $a_2(\mathcal{A}) \propto \mathcal{A}$; the quartic term $a_4(\mathcal{A})$ is taken to be constant; and higher-order terms $a_6(\mathcal{A})$ and above are neglected.

The coefficients \( a_2(\mathcal{A}) \) and \( a_4 \) are chosen to reflect the double-well structure as the RF amplitude is ramped 
\begin{align*}
    a_2(\mathcal{A}) &= 
    \begin{cases}
        \frac{\kappa_1(\mathcal{A} - \mathcal{A}_s)}{2}, & \mathcal{A} < \mathcal{A}_s, \\
        -\frac{\kappa_2(\mathcal{A} - \mathcal{A}_s)}{4}, & \mathcal{A} > \mathcal{A}_s,
    \end{cases}, \quad
    a_4 = \frac{\kappa_2}{8c^2}.
\end{align*}
This structure is derived in detail in~\cite{Wurkner2025} and guarantees a smooth transition from a harmonic trap to a double-well configuration for experimentally relevant parameter ranges.

{\textit{Appendix }} \textit{\parlabel{par:D}} {\textit{{Model calibration. }}} We calibrate the model using dynamical measurements extracted from controlled excitations of the system. Specifically, we initialize the condensate in a single well, slowly ramp to the target RF amplitude \( \mathcal{A} \), and then apply a short kick to excite sloshing modes. The resulting time evolution allows us to extract oscillation frequencies (green diamonds in Fig.~\ref{fig:1_setup}b). In regions with low barriers, both inner and outer trap frequencies can be resolved and included in the calibration. An additional set of data points comes from linear ramps and two consecutive linear ramps followed by hold times. Calibration proceeds by minimizing the \( 2 \)-norm between measured and simulated trap frequencies, using the same experimental \( \mathcal{A}(t) \) ramps, with an interior-point optimization algorithm. 

With the calibrated model, we perform optimal control by solving a boundary value problem (BVP) for the GPE using indirect methods~\cite{PhysRevA.75.023602,Mennemann_2015}. The objective is to find a time-dependent ramp \( \mathcal{A}(t) \) that transfers the initial ground state into the final double-well ground state while minimizing residual excitations. This is formulated as an energy minimization problem over the control field, with the GPE as the dynamic constraint. The accuracy of this method is critically dependent on the fidelity of the model, reinforcing the importance of robust calibration.

In Fig.~\ref{fig:5_mofel}, we present the resulting dependencies: the behavior of the model coefficients as a function of $\mathcal{A}$, the position of the potential minima, and the curvature at the bottom of the wells, all for the case of a symmetric potential.

\smallskip
{\textit{Appendix }} \textit{\parlabel{par:E}} {{OC protocol stability. }}
The data presented in the main text is obtained for an average atom number of 3000.
Robustness tests of the $OC$ protocol with respect to atom number fluctuations demonstrate reliable performance across the range of 2000–6000 atoms. We observe a degradation at lower atom numbers, which highlights the importance of nonlinear interactions ($g_\perp$) in this regime, indicating that the model is trained and optimized for dynamics governed by strong nonlinearity. Our experimental observations qualitatively agree with numerical simulations.

\end{document}